\documentclass[journal,transmag]{IEEEtran}
\usepackage{ifpdf}
\usepackage{cite}
\usepackage{color}

\usepackage{amsmath}
\usepackage{algorithmic}
\usepackage{array}
\usepackage{fixltx2e}
\usepackage{stfloats}
\usepackage{url}
\usepackage{graphicx}

\begin{document}

\title{Phase locking of spin transfer nano-oscillators using common microwave sources}

\author{\IEEEauthorblockN{R. Gopal\IEEEauthorrefmark{1},
B. Subash\IEEEauthorrefmark{2}
V. K. Chandrasekar\IEEEauthorrefmark{1}, and
M. Lakshmanan\IEEEauthorrefmark{3}}
\IEEEauthorblockA{\IEEEauthorrefmark{1}Centre for Nonlinear Science and Engineering, School of Electrical and Electronics Engineering, SASTRA Deemed University,\\ 
Thanjavur - 613 401, India}
\IEEEauthorblockA{\IEEEauthorrefmark{2} Department of Physics, Anna University, Chennai - 600 025, India }
\IEEEauthorblockA{\IEEEauthorrefmark{3} Centre for Nonlinear Dynamics, Department of Physics, 
Bharathidasan University, Tiruchirapalli - 620 024, India }}

\markboth{IEEE Transactions on Magnetics Journals}
{Shell \MakeLowercase{\textit{et al.}}: Bare Demo of IEEEtran.cls for IEEE Transactions on Magnetics Journals}

\IEEEtitleabstractindextext{%

\begin{abstract}
In this paper, we study typical nonlinear phenomenon of phase-locking or synchronization in  spin-torque nano oscillators (STNOs). To start with the oscillators are considered as uncoupled but  interlinked through either a common microwave current or a microwave field. We identify the phase locking of an array of STNOs (first for two and then for 100 oscillators) by means of injection locking which represents locking the oscillations to a common alternating spin current or a common microwave magnetic field. We characterize the locking of STNOs through both first and second harmonic lockings in an array. We find that second harmonic lockings takes  lesser value of microwave current and field when compared with the first harmonic lockings.  Our results also show that oscillating microwave current can induce integer harmonic locking while microwave field can induce both integer and several fractional harmonic lockings.  We also extend our analysis to study locking behavior of two STNOs by introducing time delay feedback and coupling through a current injection and bring out the associated locking characteristics. Finally, we have also analyzed the stability of synchronization of identical array of STNOs with current coupling by using master stability function formalism.

\end{abstract}

}

\maketitle

\IEEEdisplaynontitleabstractindextext
\IEEEpeerreviewmaketitle
\section{Introduction}

The interaction between the spins of charge carriers polarized by the pinned layer and magnetic moments in the free layer of a spin torque nano oscillator (STNO) exerts an additional torque on the magnetization of the free layer called spin transfer torque (STT). It is a newly identified physical phenomenon attracting much interest during the last decade among the scientific community\cite{Hillebrands:02}. In particular, a ferromagnetic thin film trilayer nano structure can be polarized under the action of a spin polarized current. 
In a  trilayer magnetic structure consisting of two ferromagnetic layers sandwiched by a nonferromagnetic conductor, which is a widely accepted device for spin based manipulation, the spin can be transferred from one layer to the other\cite{Stiles:06}. The phenomenon has been theoretically predicted independently by Slonczewski\cite{Slonczewski:96,tsoi,kiselev} and Berger\cite{Berger:96} through a quantum mechanical treatment of electrons flowing through a five piece nano-magnetic structure. This single composite magnetic system is capable of producing a variety of applications such as MRAMs\cite{Murugesh:09a}, microwave sources\cite{Kim:STNO,Zeng:13}, unconventional computing architecture\cite{Csaba:13}, etc., under various circumstances. Apart from digital storage applications, all the other applications utilize the self oscillating feature of the device.  The presence of spin transfer torque balances the inherent magnetic damping present in the ferromagnetic thin film thereby inducing a steady precessing state of the magnetization. As a result of giant magnetoresistance effect the magnetization oscillations lead to a high frequency current/voltage in the range of 0.1GHz to 100 GHz  and hence the device is termed as a spin transfer nano oscillator (STNO)\cite{Villard:10}. Self oscillation is a typical nonlinear phenomenon observed in many systems such as Josephson junction, relaxation oscillators, etc. The STNO is one of the smallest self oscillators known at present and the study of its nonlinear characteristics is also important from a  fundamental point of view. On the one hand, with the application of external microwave field or microwave current the various nonlinear dynamical behaviours such as bifurcations, chaos, synchronization, etc\cite{Adler:73,epl:13,pramana:15} of the STNO can be explored. On the other hand, in view of potential technology the STNO possesses various advantages such as high tunability over a wide range of frequencies, nano scale level of source, resistance to radiation damage, etc., while the deficit of the device is its low output power and spectral purity~\cite{hamdeh:14}. Both of the  later problems can be handled simultaneously by using its nonlinear character of synchronization~\cite{tiber:09,lebrun:14}.

Synchronization is a typical nonlinear phenomenon, which can be achieved through several mechanisms in coupled oscillators\cite{Pikovsky:01}.  Although there are various ways of achieving synchronization, even four hundred years ago Huygens observed the same by applying a common load to a system of two pendulum clocks. Several authors have suggested that the synchronization of STNOs coupled through various mechanisms can improve the power output of the device. Experimental observation of the external force driven synchronization in spin valve system was first reported by Rippard et al. in 2005 \cite{Rippard:l05}. The theoretical study was carried out via phase dynamics \cite{Slavin:b05, Slavin:b06}. Letting the STNOs interact themselves has also been reported as the mutual phase locking of electrically uncoupled STNOs which are in close proximity with each other\cite{kaka}.  In particular, Nakata \emph{et} al also pointed out that application of a common white noise to a system of two uncoupled STNOs acts as a medium for them to induce synchronization~\cite{nakada}. Further, the phenomenon  of synchronization/desynchronization among the magnetically uncoupled STNOs has been reported due to electrical connections\cite{Georges:08,tabor1,akerman:b12,epl:15}.  Very recently, the existence and stability of the synchronized state and the conditions to synchronize the individual precessions have also been studied in an array of $N$ serially connected identical STNOs coupled through  current~\cite{turtle}.

 However, Adler-type~\cite{Adler:73} injection locking has emerged as the  easiest method to achieve synchronization, either through an external microwave current\cite{Li:06,Georges:08a,fert:08,slavin:2009} or through a microwave magnetic field\cite{tabor:10,epl:13}. Spin wave beam\cite{Kuhn:14} produced by an external microwave field  has been shown to synchronize a record number of five STNOs\cite{spinwavebeam}.  Sychronization of spin torque oscillators results in an enhancement of the output power for tunable microwave oscillators and an increase in the quality factor of several devices~\cite{kaka,Georges:08a,mancoff:15,awad:17,tani:18}. The indication  of injection locking among driven STNOs requires a systematic study in terms of coupling and delay feedback with respect to the microwave current and field.

In this paper,  we accomplish the locking or synchronization  behaviour of STNOs  with help of  appropriate external applied oscillating sources of frequency \cite{Adler:73}.   First, we consider the example of locking in STNOs with respect to two different values of anisotropy, and then locking of STNOs is considered for $N=100$ different distributed values of anisotropy, where each STNO has been subjected to an external source in two ways (i) through an alternating current and (ii) through an external microwave magnetic field.  We report the phenomenon of first and second harmonic lockings or synchronization and find the critical strengths of the external source for the above two methods.  Fractional lockings of STNOs with the application of externally applied magnetic field has  also been identified. The micro-magnetic simulation of STNOs under microwave field is in good agreement with the experiments carried out by Urazhdin et al\cite{tabor:10}.  We have also studied the locking states of   two STNOs by introducing time delay feedback~\cite{guru:15,william:2017} and STNOs coupled through current injection~\cite{tom17}. Finally, we have also carried out a stability analysis through master stability function approach and confirmed it through numerical analysis. 

 The plan of the paper is as follows. In Sec. II, locking of a pair of STNOs by a microwave current will be discussed. Similarly, in Sec. III we discuss the locking of STNOs by an external microwave field. In Sec. IV, we point out the existence of locking regions of STNOs with respect to anisotropy field with both microwave field and current. We also examine locking behavior of STNOs through time delay feedback in Sec. V and coupling in Sec VI.  The stability analysis of synchronization of STNOs is given in Sec. VII. Finally, a critical summary of the obtained results is provided in Sec. VII.

\section{Locking of STNO by a microwave current}
\label{mic_current}
The magnetization dynamics of the free layer in the considered trilayer structure is governed by the Landau-Lifshitz-Gilbert-Slonczeski (LLGS) equation\cite{Landau:35,Laksh:12}. To numerically simulate the phenomenon of synchronization in an extended GMR trilayer structure by an ac current, we first establish a limit cycle or self sustained oscillation in the system in the absence of the ac current by solving the associated LLGS equation. On the one hand the injected direct current exerts STT (acting opposite to damping) which is used to switch the magnetization of the free layer. On the other hand, under certain circumstances the produced STT balances the inherent damping\cite{Gilbert:04} due to the effective field and makes the magnetization to oscillate in time. This typical device capable of converting the injected direct current into magnetization oscillations is termed as spin transfer/torque nano oscillator (STNO/STO).  STNO becomes a vital candidate for nano scale level source of microwaves in the telecommunication industry. Apart from technological applications, STNO is one of the smallest ever known self sustained oscillators which makes it as a strong candidate for the exploration of basic dynamics.

In our analysis, we consider two indirectly coupled STNOs with magnetization vectors represented by unit vectors with three components, ${\vec m}_{i} = (m_{i}^x,m_{i}^y, m_{i}^z)$, ${\vec m^2}_{i} = 1$, $i=1,2$. The corresponding LLGS  equation~\cite{Slonczewski:96,Berger:96} for the unit spin vectors is given by 

\begin{eqnarray}
\frac{d\vec m_{i}}{dt} & = & -\gamma\vec{m}_{i}\times
\vec{H}_{i,eff}+\lambda\vec{m}_{i}\times\frac{d\vec{m}_{i}}{dt} 
\nonumber \\ & &-\gamma \tilde a_i(t)\vec{m}_{i}\times(\vec{m}_{i}\times\hat{m}_{p}),~~~ i=1,2, \label{ccoupled}
\end{eqnarray}
where the spin current densities $\tilde a_{i}(t)$ are given by 
\begin{align}
\tilde a_{i}(t)= \frac{\hbar \eta}{2 m_0 V e}[I_{i,dc} + I_{ac} \cos (\Omega t)].  
\label{ccoupleda}
\end{align}
Here $\gamma$ is the gyromagnetic ratio, $\lambda$ is the damping coefficient, $\eta$ is the spin polarization ratio, $V$ is the volume of the free layer and $m_{0}$ is the saturation magnetization, $I_{i,dc}$, $i=1,2$ is the bias current for the $i^{th}$ oscillator and $I_{ac}$ is the common oscillating current with frequency $\Omega$ supplied to the STNOs.
Further, we also assume the unit vector defining the polarization of the spin current $\hat{m}_p=(1,0,0)$ and that the effective fields are
\begin{align}
\vec{H}_{i, eff} &=& \vec{H}_{exchange}+\vec{H}_{applied}+\vec{H}_{i,anisotrophy}+
\nonumber\\&&\vec{H}_{i,demag}, i=1,2. 
\label{magfield}
\end{align}
Considering homogeneous magnetization in the ferromagnetic free layer, we assume that $H_{exchange} = 0$. We assume an easy plane anisotropy, $\vec{H}_{i,anisotropy} = (\kappa_{i} m_{i}^x , 0, 0)$, where $\kappa_{i}$ is the strength of the anisotropy. We also assume a demagnetization field $\vec{H}_{i,demag} = -4\pi m_0 (0, 0, m_{i}^z)$, and for the present geometry $m_{0}$ is the saturation magnetization, and the external field $\vec{H}_{applied} = (h_{dc}, 0, 0)$ is applied along the $x$-axis. The effective magnetic fields of the two STNOs (\ref{magfield}) are then given by
\begin{align}
\vec{H}_{i, eff} & = h_{dc}   \hat i + \kappa_{i} m_{i}^x  \hat i - 4 \pi m_0 m_{i}^z \hat k, i=1,2.  
\label{hcoupledheff}
\end{align}

To start with, in our study, we consider an array of two STNOs (\ref{ccoupled}) and consider their synchronization. This is further extended to larger arrays later on. In the case of the two STNOs differing by their anisotropy strengths, the later are chosen as $\kappa_1 = 40$ Oe and $\kappa_2 = 45$ Oe. The two STNOs are magnetically uncoupled, that is their magnetizations  have no interactions such as dipolar interaction, etc. between their magnetizations. They are also provided with the currents  $I_{i,dc}+I_{ac}cos(\Omega t), i=1,2$,  where the bias current $I_{i,dc}, i=1,2$ is separately given to the two STNOs, such that they are also electrically uncoupled. Like in the case of Huygens' pendulum, we assume a common oscillating current of strength $I_{ac}$, having a frequency $\Omega$, which is supplied simultaneously to the STNOs. 

 The material parameters corresponding  to the typical permalloy (NiFe) are given as follows.
 $\gamma=0.01767$ O$e^{-1}$n$s^{-1}$, $\lambda=0.02$, $\eta=0.35$, $4\pi m_{0}=8.4k$Oe and $V=100 nm \times 200 nm \times 5 nm$ is the volume of the free layer. We further consider the bias current $I_{i,dc} =11.5 $mA and the strength of the static magnetic field $h_{dc} = 200$ Oe.  In the numerical simulations, initially starting from an arbitrary point other than the stable point or saddle point on the sphere $|\vec m| =1$  and by applying Runge-Kutta-Cash-Karp algorithm, the limit cycle oscillation is usually achieved after several nanoseconds. Once the self sustained magnetization oscillation becomes stable, we determine its natural frequency by measuring the inverse of the  average of the time intervals between successive minima/maxima of the time series of $m_{i}^z$. 

       Before applying the ac current (i.e. $I_{ac}=0$ in (\ref{ccoupled})), we measure the frequency of the two STNOs to be 11.11 and 10.92 GHz, $\kappa_1 = 40$ Oe and $\kappa_2 = 45$ Oe respectively, where they are dynamically possessing limit cycle oscillations. Now we turn on the ac current, $I_{ac} > 0$, which induces an ac component in the spin transfer torque strength. The frequency of the ac current is chosen to be close to the measured natural frequency of the individual STNOs, for instance we choose $\Omega = 10 $GHz. In order to achieve  locking we gradually increase the amplitude/strength of the ac component $I_{ac}$  from $I_{ac} = 0$ in (\ref{ccoupled}).  
\begin{figure}
\begin{center}
\includegraphics[width=1.0\columnwidth]{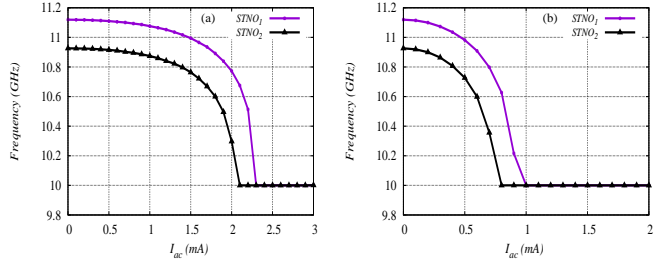} 
\caption{(a) Harmonic locking: Frequencies of the two STNOs are plotted by increasing the strength of the microwave current($I_{ac}$) whose frequency is $\Omega = 10$ GHz. (b) Second harmonic locking: Frequencies of the two STNOs are plotted by increasing the strength of the microwave current whose frequency $\Omega = 20$GHz.}
\label{iacw10}
\end{center}
\end{figure}

\begin{figure}
\begin{center}
\includegraphics[width=1.0\columnwidth]{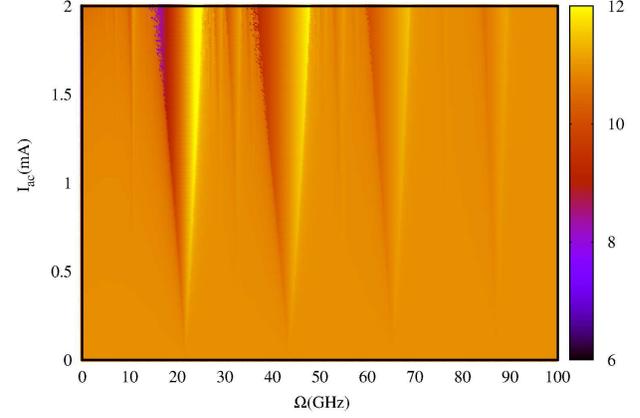} 
\caption{Locking characteristics of the STNO under microwave current: Arnold tongue diagram is shown by plotting the frequency of $m_{1}^{z}$ against external microwave current frequency ($\Omega$) and its strength($I_{ac}$). The gradient areas in the Arnold tongue diagram represent the "locked" state regions of the oscillator.}
\label{iacat}
\end{center}
\end{figure}

We have analyzed the synchronization dynamics of the two STNOs and presented the relevant results in Fig.\ref{iacw10}(a). Note that the second STNO which has the frequency nearer to the frequency of the microwave current gets locked first and it has the lesser critical strength $I_{ac} = 2.1 $mA to achieve it. On further increasing the strength of the ac current, both the oscillators get locked to the external source for $I_{ac} = 2.3 $mA, which may be noted as the critical strength for the  frequency locking of the above two STNOs.   Here the phases of the two oscillators become locked to the phase of the driving signal, with a specific relationship, that is 1:1 between them. So this state of locking is generally called as first-harmonic locking.  One can also observe the higher harmonic locking by increasing the frequency of the ac current. 

As an example, we have plotted the  locking dynamics of two STNOs driven by the ac current of frequency $\Omega = 20$GHz, see Fig.\ref{iacw10}(b). Similar to the case of $\Omega = 10$GHz, initially when $I_{ac}=0$ the two STNOs  exhibit limit cycle oscillations. On increasing the strength of the ac current, the second oscillator gets synchronized/locked first and on further increasing the value of $I_{ac}$ the first oscillator also gets locked when the strength of the ac current reaches 1~mA. One may note that both the oscillators get locked to the frequency 10 Ghz which is $\Omega/2$ and hence this frequency locked state is called as  second harmonic locking.  One can also observe that locking of STNOs can be achieved for even lesser values of $I_{ac}$ in the case of second harmonic locking when compared with the first harmonic locking.

Higher harmonics can be found on further increasing the external current frequency. 
It is not necessary that the ac frequency $\Omega$ is close to the natural frequency of spin valve oscillators($f$). Whenever $\Omega/f$ is rational, one may have an opportunity to observe the synchronization phenomenon. Suppose  n$\Omega$ is close to m$f$, where both m and n are positive integers, the response oscillator may oscillate exactly n periods when the driven force oscillates m periods\cite{Balanov:08,Yang:07}. Thus, one observes the relation
\begin{align}
\mbox{n}~T_{stno} =  \mbox{m}~T_{ac}
\end{align}
or
\begin{align}
\mbox{m}~f =\mbox{n}~\Omega,
\end{align}
where the $T_{stno}$ and $f$ are the period and frequency of the spin transfer oscillator in response to the driven ac current, $T_{ac}$ and $\Omega$ are the period and frequency of the ac current. In order to explore the full locking dynamics of the STNOs with an ac current, we need to vary the ac frequency $\Omega$. 

In Fig.\ref{iacat}, we have plotted the frequency of the single STNO against the parameters of the ac frequency $\Omega$ and ac strength $I_{ac}$. The gradient regions are the regions of locked states of the STNO with the ac current.  The picture describes the locking zone in relation with the coupling of external force. This characteristic is referred to as Arnold's tongue and the corresponding picture as the Arnold tongue diagram\cite{Pikovsky:01}.  In the first tongue like region, for a fixed ac frequency, say $\Omega=22$GHz, by  tracing the Fig.\ref{iacat} upwards for the increasing values of $I_{ac}$, one can find the second harmonic locking ($f=\Omega/2$) of STNO with that of the ac current. Similarly  the third harmonic locking ($f=\Omega/3$) around $\Omega=33 $GHz, fourth harmonic locking($f=\Omega/4$) around $\Omega=44 $GHz and higher harmonics can be seen in Fig.\ref{iacat}. For a constant strength of the microwave current (i.e. tracing the Fig.\ref{iacat} horizontally) the thickness of the Arnold tongue represents the locking range/bandwidth of the oscillator. One may also observe that the thickness of the gradient/tongue or locking range is increasing linearly with the strength of the ac current.

\section{Locking of STNO by an external microwave field}
Now we switch off the ac current $I_{ac} = 0$ in (\ref{ccoupleda}) and apply a common microwave magnetic field simultaneously along the two STNOs in the form $\vec{H}_{applied} = (h_{dc}+h_{ac} cos\omega t, 0, 0)$ in (\ref{magfield}). The effective fields of the two STNOs are given by
\begin{align}
\vec{H}_{i, eff} = (h_{dc} + h_{ac} \cos \omega t) \hat i + \kappa_{i} m_{i}^x  \hat i - 4 \pi m_0 m_{i}^z \hat k, ~ i=1,2\nonumber \\ 
\label{hcoupledheff}
\end{align}
Note that here the two STNOs are driven by the same magnetic field having both dc and ac components of strength $h_{dc}$ and $h_{ac}$, respectively. Here $\omega$ is the frequency of the ac magnetic field and all the remaining parameters are same as in the previous section  (Sec.\ref{mic_current}). We consider the bias current $I_{i,dc} =11.5 $ mA and the strength of the static magnetic field $h_{dc} = 200$ Oe.   
\begin{figure}
\begin{center}
\includegraphics[width=1.0\columnwidth]{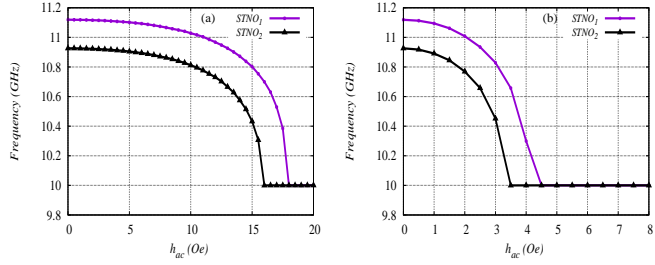} 
\caption{(a) Harmonic locking: Frequencies of the two STNOs are plotted by increasing the strength of the microwave magnetic field of frequency $\omega = 10$GHz. (b) Second harmonic locking: Frequencies of the two STNOs are plotted by increasing the strength of the microwave magnetic field frequency $\omega = 20$GHz.}
\label{hacw10}
\end{center}
\end{figure}

\begin{figure}
\begin{center}
\includegraphics[width=1.0\columnwidth]{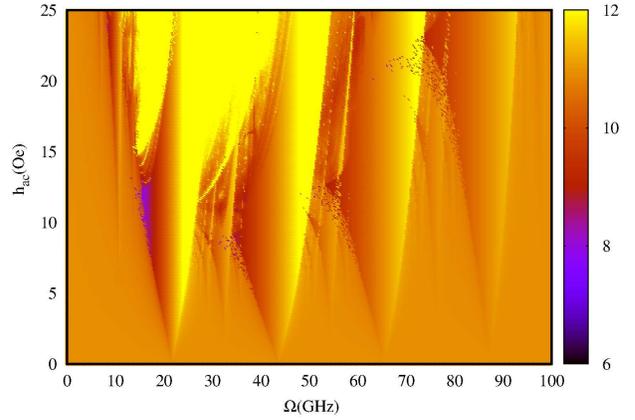} 
\caption{Locking characteristics of the STNO by microwave field: Arnold tongue diagram is shown by plotting the frequency of $m_{1}^{z}$ against external microwave magnetic field frequency ($\omega$) and its strength($h_{ac}$). The gradient areas of the Arnold tongue diagram represent the "locked" states regions of the oscillator. The distorted yellow regimes between the tongues are due to the overlap of synchronization tongues resulting in multi-stability and hysteresis.}
\label{hacat}
\end{center}
\end{figure}

Before applying the ac field (i.e. $h_{ac}=0$ in Eq.(\ref{hcoupledheff})), we measure the frequency of the two STNOs to be 11.11 and 10.92 GHz, respectively. (Since we assume that all the parameters as identical to the previous section, the free running or limit cycle frequencies of the two STNOs remain unchanged.) Now we turn on the ac field, which induces an ac component in the applied external magnetic field as in Eq.(\ref{hcoupledheff}). The frequency of the ac field is chosen to be close to the measured natural frequency of the individual STNOs, for instance we choose $\omega = 10 $GHz. In order to achieve synchronization we gradually increase the amplitude/strength of the ac component $h_{ac}$  from $h_{ac} = 0$.

We have portrayed the synchronization dynamics of the two STNOs  in Fig.\ref{hacw10}(a). The solid red line and dotted green line correspond to the frequencies of the first and second STNOs, respectively, calculated for the increasing values of amplitude ($h_{ac}$) of the ac field. Similar to the previous case, the second STNO which has a frequency nearer to the frequency of the microwave field gets locked first and it has the lesser critical strength $h_{ac} = 16 $ Oe to achieve it.  On further increasing the strength of the ac field, both the oscillators get locked to the external source for $h_{ac} = 18 $ Oe, which may be noted as the critical strength  of the ac field required for the synchronization of the above two STNOs.   The STNOs are exhibiting first harmonic locking, since the frequency  of the  oscillators become locked to the frequency of the driving signal with a relationship 1:1 between them.  

In Fig.\ref{hacw10}(b), we have shown the second harmonic locking of STNOs with the source. Here the frequency of the microwave field is chosen as $\omega = 20$ GHz. The two oscillators are locked to the frequency 10 GHz and hence having a relationship of 2:1 with the driving source.

\begin{figure}
\begin{center}
\includegraphics[width=1.0\columnwidth]{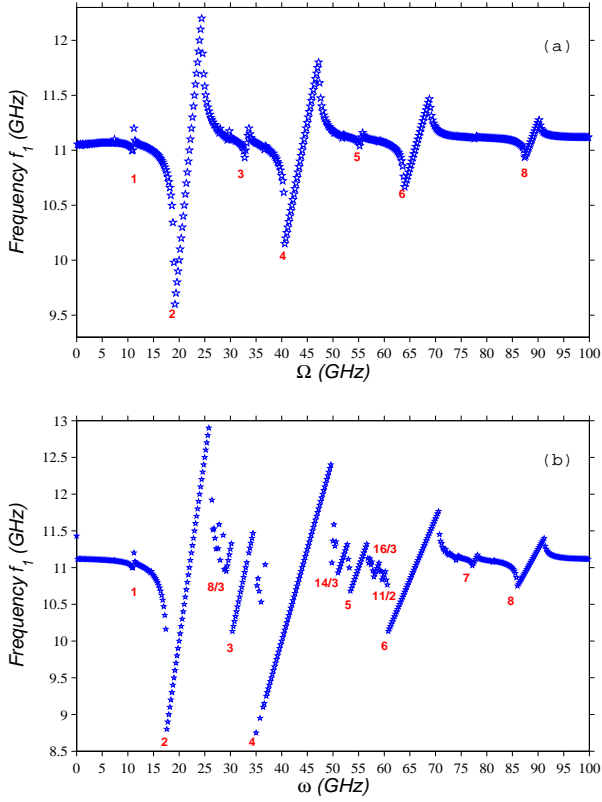} 
\caption{Locking characteristics : (a)The frequency of the first STNO is plotted against the frequency of the external microwave current ($\Omega$)  for a particular strength of $I_{ac} = 1.22 $ mA. It shows the various harmonic lockings of STNO with the external source. (b) The frequency of the first STNO is plotted against the frequency of the external microwave field ($\omega$) for a fixed value of the strength of the microwave magnetic field of strength $h_{ac} = 8.6 $ Oe. It shows the various integer and fractional lockings of STNO with the external microwave field.}
\label{flock}
\end{center}
\end{figure}

Fig.\ref{hacat} depicts the Arnold-tongue/frequency characteristic of the first STNO with the driving microwave field. As in the previous case, STNOs are undergoing integer harmonic locking with the source. Now in the present case, by the application of the microwave magnetic field, we find that in addition to integer harmonic lockings there are several fractional lockings/synchronization in the system, that is fractional ratios between the oscillator and the microwave field frequencies.  As can be observed, the fractional locking is achieved for several non-integer values. From a theoretical point of view, it has been demonstrated that to observe fractional synchronization the microwave field should be applied to the free layer in order to break the symmetry of the system~\cite{tabor:10,tabor1:10,Zeng:13}. The distorted yellow regimes between the tongues are due to the overlap of synchronization tongues resulting in multi-stability and hysteresis~\cite{Balanov:08}. In addition chaos can occur which leads to the destruction of the onset of synchronization.

\begin{figure}
\begin{center}
\includegraphics[width=0.95\columnwidth]{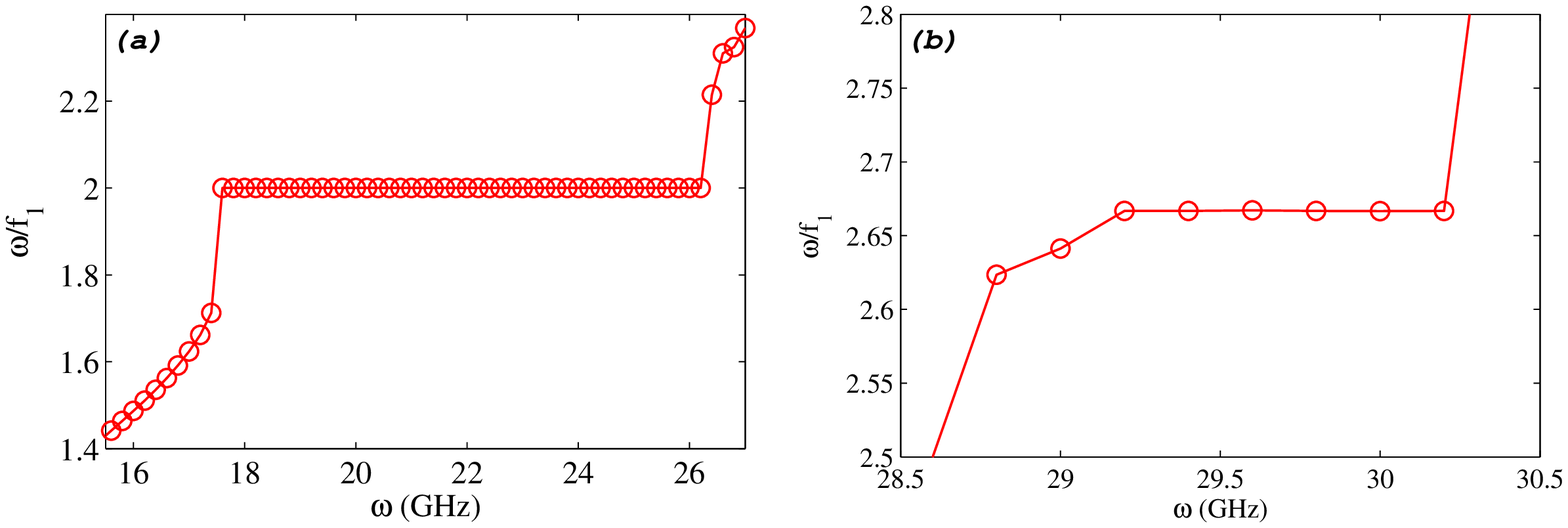}
\caption{Locking characteristics of the STNO are shown by plotting the ratio $\omega/f_1$ against the external source frequecy $\omega$ for a fixed value of the strength $h_{ac} = 8.6 $ Oe.  (a) Second harmonic locking  ($\omega/f_1$=2) and (b) Fractional locking ($\omega/f_1 = 2.66 = 8/3$) are shown for different ranges of $\omega$.}
\label{rlock}
\end{center}
\end{figure}

In Fig. \ref{flock}(a), we have shown the locking characteristics of the two STNOs under the forcing of common microwave current of strength $I_{ac}= 1.22$ mA. One may observe that the first STNO is undergoing various integer harmonic lockings with the external current. Fig. \ref{flock}(b) elucidates the locking characteristics of the first STNO with the external microwave field source. By varying the frequency ($\omega$ GHz) of the external microwave field source, the frequency of the first STNO is plotted. The frequecy of the STNO remains close to the natural frequency (i.e. $f_1$ when $h_{ac}=0$) 11.1 GHz at most of the ranges. In addition there are ranges of $\omega$ where the quantity $\omega/f_1$ take integer values say 1,2,3,..8. Apart from integer relations there are several rational relations between the STNOs and the driven microwave field, for instance 8/3,14/3,16/3 and 11/2. The values of $\omega/f_1$ are written below the corresponding locking regimes.  Further, the locking characteristics of the STNOs are also plotted with respect to the ratio of the quantity $\omega/f_1$ versus external source frequency $\omega$. Figs. \ref{rlock}(a) and \ref{rlock}(b) depict the second harmonic locking and fractional lockings with respect to the external source frequency. We wish to point out here such integer and fractional lockings have been experimentally identified in micromagentic experimental studies by  Urazhdin et al~\cite{tabor:10}.

\section{Locking characteristics of STNO with respect to anisotropy}

\begin{figure}
\begin{center}
\includegraphics[width=1.0\columnwidth]{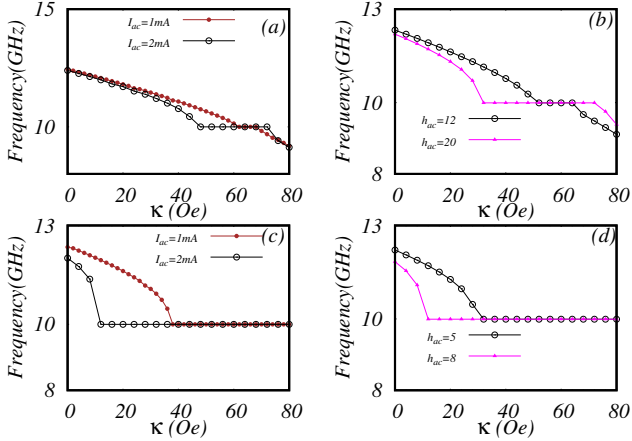} 
\caption{Harmonic locking: Frequencies of the two STNOs are plotted by increasing the strength of the anisotropy  as a function of (a) microwave current $I_{ac}$, (b) microwave field $h_{ac}$ whose frequency is $10$ GHz. Second Harmonic locking: Frequencies of the two STNOs are plotted by increasing the strength of the anisotropy  as a function (c) microwave current $I_{ac}$, (d) microwave field $h_{ac}$ whose frequency is $20$ GHz.}
\label{fig4}
\end{center}
\end{figure}

  In order to understand how the locking characteristics of STNOs vary with respect to the influence of the anisotropy, we first make the choice $\kappa_{i}= \kappa$, and plot the frequency versus anisotropy ($\kappa$) curve for different values of $I_{ac}$ (Fig.\ref{fig4}(a) and Fig.\ref{fig4}(c)) and $h_{ac}$ (Fig.\ref{fig4}(b) and Fig.\ref{fig4}(d)) for the first harmonic and second harmonic with  values of  frequency $10$GHz and $20$GHz respectively. From Fig.\ref{fig4}, we have identified that the locking region of STNO is enhanced for second harmonic when compared with the first harmonic for both applied current $I_{ac}$ and magnetic field $h_{ac}$. Further one may also note that on increasing the value of $I_{ac}$ and $h_{ac}$, the locking region also increases for both first and second harmonics.

\begin{figure}
\begin{center}
\includegraphics[width=1\columnwidth]{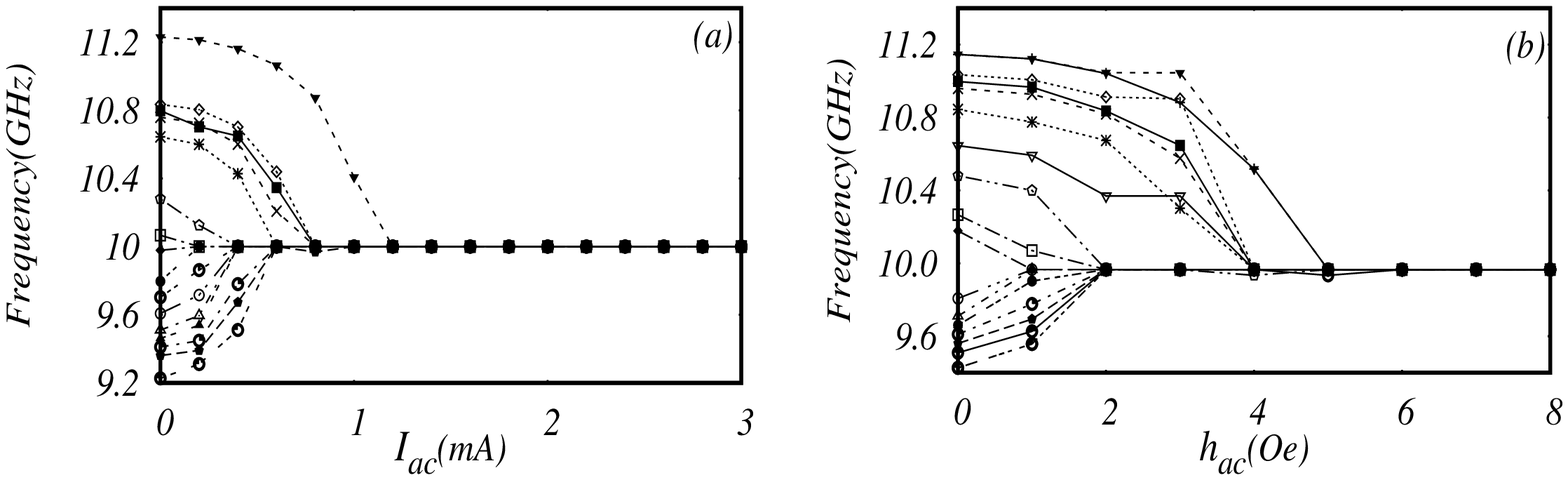} 
\caption{Second harmonic locking: Frequencies of the selected number of ($18$) STNOs are plotted by increasing the  (a) microwave current $I_{ac}$, (b) microwave field $h_{ac}$ whose frequency is $20$ GHz. Here, the value of anisotropy is distributed in the range $\kappa\in(30,80)$ for $N=100$ number of oscillators.}
\label{fig8}
\end{center}
\end{figure}

One can also note that in Fig.\ref{fig4}(c) and \ref{fig4}(d), the second harmonic lockings occur in the  range $\kappa \in(20,80)$ for $I_{ac}=2$mA and $h_{ac}=8~$Oe. This study indicates that for the value of frequency  $20$GHz with the value $I_{ac}=2$mA or $h_{ac}=8~$Oe, one is able to achieve synchronization for $N=100$ number of STNOs once the value of $\kappa$ is distributed between 20 and 80. Therefore, we have also extended our analysis to large number of oscillators. We have verified the above locking features for $N=100$ in Figs.\ref{fig8}(a)-(b) where we have depicted the locking characteristics of STNOs as a function of ac current and microwave field for the choice of anisotropy uniformly distributed between 30 to 80.

\begin{figure}
\begin{center}
\includegraphics[width=1\columnwidth]{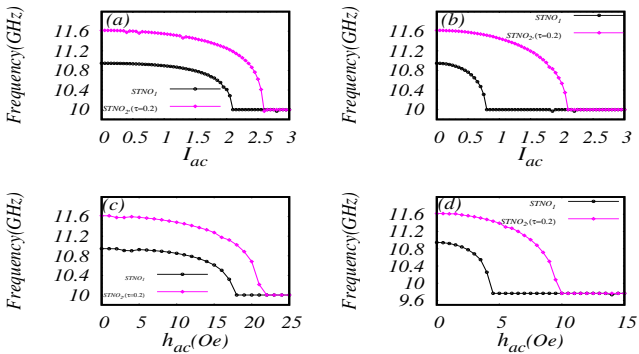} 
\caption{Frequencies of the two STNOs are plotted by increasing the strength of the microwave current($I_{ac}$)  whose frequency is (a) $\Omega = 10$ GHz. (b) $\Omega = 20$GHz;Frequencies of the two STNOs are plotted by increasing the strength of the microwave field ($h_{ac}$) whose frequency is (c) $\omega = 10$ GHz. (d) $\omega = 20$GHz.}
\label{delay1}
\end{center}
\end{figure}

\section{Effect of time delay}
In order to study the effect of time delayed feedback on STNO, we consider the spin current densities $\tilde a_1(t)$ and $\tilde a_2(t)$ in Eq.(\ref{ccoupled})  which are taken to be time dependent with or without self delay feedback,
\begin{align}
\tilde a_1(t) & = a(t)&,  \nonumber \\ \;
\tilde a_2(t) & = a(t) (1+\Delta j m^{z}_{2}(t-\tau)). 
\end{align}
where $a(t) = \frac{\hbar \eta}{2 m_0 V e}[I_{dc} + I_{ac} \cos (\Omega t)]$ and the expression $(t-\tau)$ represents the time shift by $\tau (ns)$. Here, the output of the spin torque nano oscillator ${m}^{z}_{2}$ is typically considered by means of delay feedback $\tau$  with proper gain factor $\Delta j=0.1$~\cite{william:2017}. Very recently, delayed feedback in a  spin torque nano oscillator has been shown to lead to various types of dynamical states, where transitions between different oscillation modes occur by using large delay and feedback amplitude due to the strongly nonlinear nature of magnetization dynamics~\cite{william:2017}. Hence, we choose minimum delay feedback strength in our studies.

\begin{figure}
\begin{center}
\includegraphics[width=1.0\columnwidth]{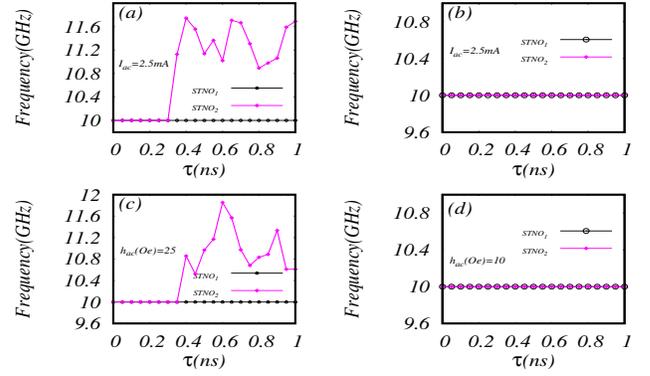} 
\caption{Frequencies of the two STNOs are plotted by increasing the delayed feedback value of oscillator 2 with microwave current  whose frequency is (a) $\Omega = 10$ GHz (b) $\Omega = 20$GHz; Frequencies of the two STNOs are plotted by increasing the delayed feedback of as a function of microwave field by whose frequency is (c) $\omega = 10$ GHz (b) $\omega = 20$GHz}
\label{delay2}
\end{center}
\end{figure}

 In Fig.~\ref{delay1}, we show the locking behavior of the oscillators by introducing delay feedback in the second  STNO as a function of microwave current $I_{ac}$ and microwave field $h_{ac}$ in terms of first and second harmonic frequencies. The variation of the frequencies of oscillator 1  and oscillator 2 ($\tau=0.2$) with increasing microwave current is shown in Figs~\ref{delay1}(a)-(b). The figures show that the delayed feedback oscillator 2 takes large current in order to synchronize with the oscillator 1 as in the case of second harmonic locking (see Fig~\ref{delay1}(b)). Similar results have also been obtained in Fig.~\ref{delay1}(d), where one can observe that large microwave field is needed (see also Fig. ~\ref{delay1}(c)) in order to get locking behavior for delayed feedback oscillator 2 with oscillator 1 for the case of second harmonic locking. Further, we also analyzed the locking behavior by inducing delay feedback with fixed values of microwave current $I_{ac}$ and microwave field $h_{ac}$ for both first and second harmonic lockings in Fig.~\ref{delay2}. We first consider the curves of Fig.~\ref{delay2}(a) and Fig.~\ref{delay2}(c) corresponding to first harmonic locking with fixed values of $I_{ac}$ and $h_{ac}$. They show that for certain low values of optimal delay, self-delayed feedback oscillator 2 exhibits locking behavior with oscillator 1, and once the delay value exceeds certain critical value the locking behavior completely vanishes and the frequency of the delayed feedback oscillator gets enhanced randomly. However, we find that the delayed feedback oscillator 2 is always locked with the oscillator 1 by increasing the delay as in the case of second harmonic locking as clearly illustrated in Figs.~\ref{delay2}(b) and (d).

\section{Effect of coupling}
In this section, we focus on the effect of coupling in the locking behavior of the spin torque nano-oscillators through the current injection. This study represents STNOs coupled through a current injection which denotes current ejected from one layer is injected into the other layer in the presence of either microwave current or microwave magnetic field. Usually, this type of current injection excites additional spin torque in the magnetization of the layer and its also depends upon the direction at which the influence takes place from one layer to another~\cite{tom17}.

In order to study the effect of coupling, the spin current densities $\tilde a_{1}$ and $\tilde a_{2}$ in Eq.(\ref{ccoupled}) can be taken as 
$\tilde a_1=a+a^{coup}_{1}$ and $\tilde a_2=a+a^{coup}_{2}$ respectively. The  terms  $a^{coup}_{1}$, $a^{coup}_{2}$ here denotes the injection of current from one layer to another and they are given by,\\
\begin{align}
a^{coup}_{1}=a\epsilon[m^{x}_{1}(t)-m^{x}_{2}(t)], \nonumber \\
a^{coup}_{2}=a\epsilon[m^{x}_{2}(t)-m^{x}_{1}(t)]. \nonumber 
\end{align}
Here, $\epsilon$ denote the strength of the coupling and $a$ denotes value spin current density(\ref{ccoupleda}).

\begin{figure}
\begin{center}
\includegraphics[width=1\columnwidth]{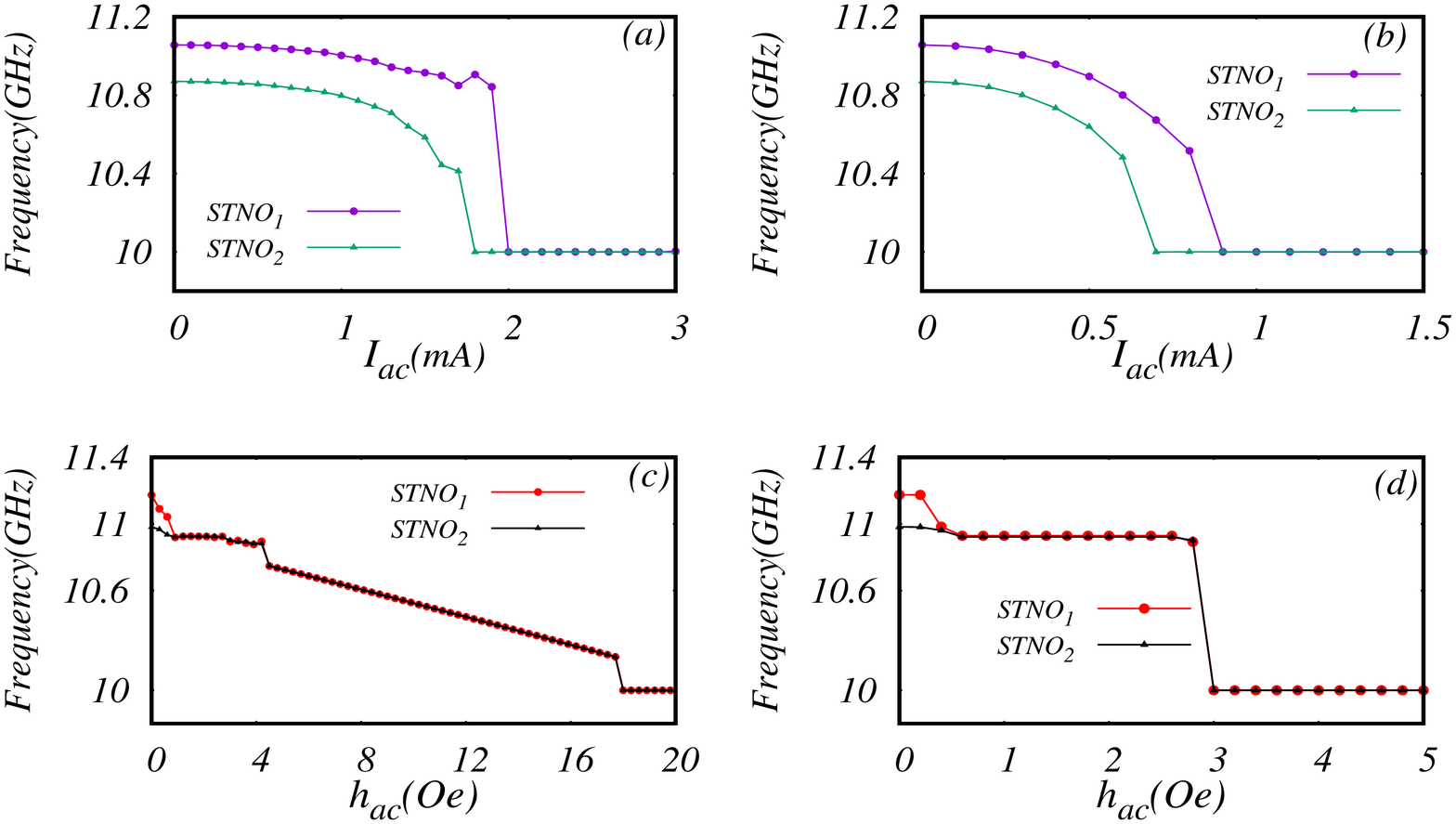} 
\caption{Frequencies of the two STNOs on increasing the values of $I_{ac}$ in the presence of current coupling strength $\epsilon=0.3$ with frequency values of  (a) $\Omega = 10$ GHz (b) $\Omega = 20$GHz; Frequencies of the two STNOs plotted by increasing the values microwave field in presence of coupling strength $\epsilon=0.5$ with frequency of (c) $\omega = 10$ GHz (d) $\omega = 20$GHz}
\label{coup}
\end{center}
\end{figure}

In Fig.~\ref{coup} we have plotted the locking of  coupled STNOs in the presence of coupling strength $\epsilon$ in the cases of both microwave current and microwave field. 
The locking behavior of STNOs as a function of $I_{ac}$ in the presence of $\epsilon$ shows that the frequency locking with external source is attained for advanced values of $I_{ac}$ in the  cases of both first and second harmonic lockings when compared with absence of coupling (See Figs.\ref{coup}(a)-(b)). The coupling can take minimum values of current to produce the locking of coupled STNOs with external microwave source.

Similarly, the frequency locking behavior is also  studied with respect to the microwave field. When  the frequency of the common microwave field is fixed as $\omega=10$GHz and by varying the values of $h_{ac}$ due to the presence of coupling $\epsilon$, the two STNOs are seen to exhibit locking behavior when $h_{ac} \geq 0.7$.  Further different fractional lockings  are seen to exist  with increasing $h_{ac}$, and finally $(1:1)$ locking is attained with external source at a value of $h_{ac}=17.8$Oe. Then, when the value of $\omega$ is fixed as $20$GHz, constant fractional locking arises when $h_{ac}$ takes  values between $0.6<h_{ac}<2.8$, and suddenly $2:1$ frequency is attained once the values of $h_{ac} \geq 2.8$. The above study helps to tune the desired frequency locking among the STNOs with external source in the presence of coupling. 

\section{Stability analysis of Synchronization of STNOs}
In this section, we summarize our detailed analysis of how one can obtain stability of synchronization of large arrays of identical STNOs through the analysis of  master stability function formalism~\cite{pecora}. In this regard under a stereographic projection~\cite{lak} Eq.(\ref{ccoupled}) can be equivalently rewritten using the complex scalar function
\begin{align}
 z_{j}(t)=\frac{m_{j}^{x}+im_{j}^{y}}{1+m_{j}^{z}} \nonumber
\end{align} 
or equivalently
\begin{align}
m_{j}^{x}=\frac{z_{j}+\bar{z}_{j}}{1+|z_{j}|^{2}};\;\; m_{j}^{y}=\frac{-z_{j}-\bar{z}_{j}}{1+|z_{j}|^{2}};\;\; m_{j}^{z}=\frac{1-|z_{j}|^{2}}{1+|z_{j}|^{2}} \nonumber
\end{align}  
as 
\begin{align}
\frac{1+\lambda^{2}}{\gamma}\frac{dz_{j}}{dt}=(1+i\lambda)\Bigg\{\frac{-i(1-z_{j}^{2})}{2}\Bigg[h+\frac{\kappa(z_{j}+\bar{z}_{j})}{1+|z_{j}|^{2}}\Bigg] \nonumber \\
-i4\pi m_{0}z_{j}\frac{(1-|z_{j}|^{2})}{1+|z_{j}|^{2}}+\frac{\tilde a_{j}}{2}(1-z_{j}^{2}) \Bigg \}  
\label{complex} 
\end{align}
where $h=h_{dc}+h_{ac}cos(\omega t)$, For the above array of STNOs coupling is considered through spin current density $\tilde a_{j}=a(t)(1+\frac{\epsilon}{N}\sum_{k=1}^{k=N}(m^{x}_{k}(t)-m^{x}_{j}(t)))$ and $a(t)=\frac{\hbar \eta}{2 m_0 V e}[I_{dc} + I_{ac} \cos (\Omega t)]$.

Rewriting the above equation in terms of real and imaginary variables $z_{j}=x_{j}+iy_{j}$, we get
\begin{eqnarray}
&&\frac{\gamma}{1+\lambda^{2}}\dot{x}_{j}=f(x_{j},y_{j})+\frac{a\epsilon}{N}\sum\Bigg[R_{k}(x_{k},y_{k})\nonumber \\ 
&&\qquad-R_{j}(x_{j},y_{j})\Bigg] \Bigg[\frac{1-(x_{j}^{2}+y_{j}^{2})}{2}+
\alpha x_{j}y_{j}\Bigg] \nonumber\\
&&\frac{\gamma}{1+\lambda^{2}}\dot{y}_{j}=g(x_{j},y_{j})+\frac{a\epsilon}{N}\sum\Bigg[R_{k}(x_{k},y_{k})\nonumber \\ 
&&\qquad-R_{j}(x_{j},y_{j})\Bigg] \Bigg[\alpha (\frac{1-(x_{j}^{2}+y_{j}^{2})}{2})-
x_{j}y_{j}\Bigg] \nonumber, 
\end{eqnarray}

where
\begin{eqnarray}
f(x,y)&=&\Bigg\{(\frac{\alpha h}{2}+\frac{a}{2})(1-r_{1}^{2}))+(\alpha \kappa+4\pi m_{0}\alpha )x G(x,y)\nonumber \\
&+&(a\alpha-h)xy-\frac{2\kappa x^{2}y}{1+r_{2}^{2}}+4\pi_{0}yG(x,y) \Bigg \} \nonumber\\
g(x,y)&=&\Bigg\{(\frac{a\alpha}{2}-\frac{h}{2})(1-r_{1}^{2})-(\kappa+4\pi m_{0})xG(x,y)\nonumber \\ 
&-&(a+\alpha h)xy-\frac{2\alpha  \kappa x^{2}y}{1+r_{2}^{2}}+4\pi_{0}\alpha yG(x,y)\Bigg \} \nonumber
\label{complex1},
\end{eqnarray}
\begin{align}
&& R(x,y)=\frac{2x}{1+x^{2}+y^{2}}, G(x,y)=\frac{(1-r^{2})}{(1+r^{2})}, \nonumber
\end{align}
and
\begin{align}
r_{1}^{2}=x^{2}-y^{2}, r_{2}^{2}=x^{2}+y^{2}. \nonumber
\end{align}
Further, the synchronized  behavior of large arrays of identical STNOs and their stability properties can be easily be identified by finding the transverse Lyapunov exponents associated with the above system of equations. In the synchronized manifold $x_{1}=x_{2}=...=x_{N}=x$ and $y_{1}=y_{2}=...=y_{N}=y$, the set of variational equations of the above system can be written as 
\begin{align}
\frac{\gamma}{1+\lambda^{2}} \dot{u}_{j}&=&f_{x}(x,y)u_{j}+f_{y}(x,y)v_{j} 
+\mu_{j} a\epsilon (R_{x}u_{j}+R_{y}v_{j})\nonumber\\
&&\times \Bigg (\frac{(1-r_{1}^{2})}{2}+ \alpha xy \Bigg),
\label{msf1}
\end{align}
\begin{align}
\frac{\gamma}{1+\lambda^{2}} \dot{v}_{j}&=&g_{x}(x,y)u_{j}+g_{y}(x,y)v_{j}+ 
\mu_{j} a\epsilon (R_{x}u_{j}+R_{y}v_{j})\nonumber\\&&\times \Bigg (\frac{\alpha(1-r_{1}^{2})}{2}-xy \Bigg ),  
\label{msf2}
\end{align}

 where  $u_{j}=\boldsymbol {\delta^x \xi_j}$ and  $v_{j}=\boldsymbol {\delta^y \xi_{j}}$ denote the perturbation from the synchronized manifold.  Here $\boldsymbol{{\delta^{x}}}=(\delta_1^x,\delta_2^x,...,\delta_N^x)$ and $\boldsymbol{{\delta^{y}}}=(\delta_1^y,\delta_2^y,...,\delta_N^y)$.  $ \delta_j^x$ and $\delta_j^y$ are the deviation of $x_j$ and $y_j$ from the  synchronized solution $(x, y)$.
   
\begin{figure}
\begin{center}
\includegraphics[width=1.0\columnwidth]{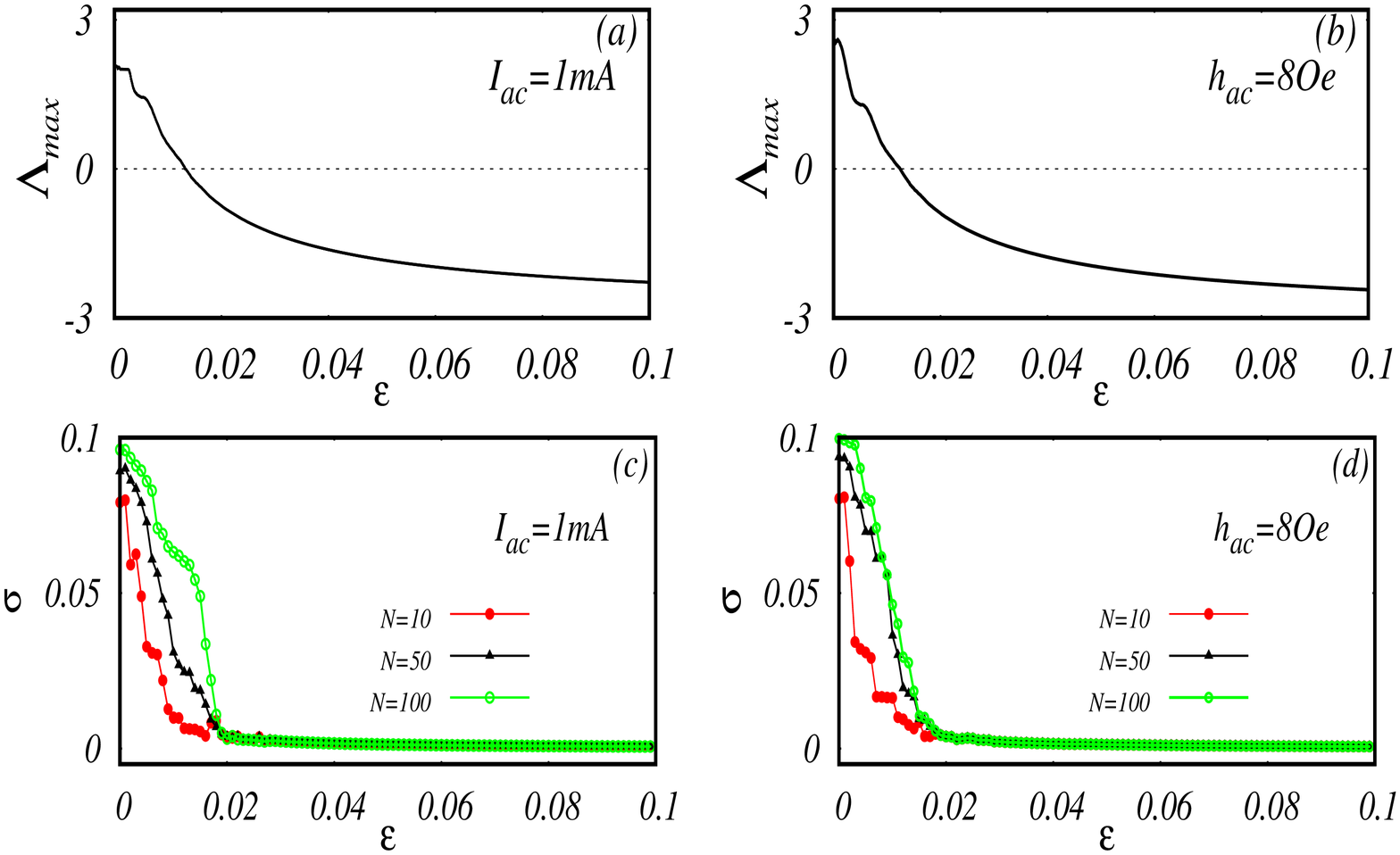} 
\caption{(a)-(b)~Largest Lyapunov exponent $\Lambda_{max}$ and (c)-(d)~Standard deviation $\sigma$ versus coupling strength $\epsilon$ for fixed values of microwave current $I_{ac}$ and microwave field $h_{ac}$. The values of anisotrophy is $\kappa=45 Oe$. Other parameters are as in Fig.\ref{iacw10} }
\label{fig12}
\end{center}
\end{figure}

In the system (\ref{msf1}) and (\ref{msf2}) $\boldsymbol{\xi_j}$ is the  eigen vector of the global coupling matrix corresponding to the eigen values
 $\mu_1=0$, $\mu_j=-1$, $j=2,3,...N$.  The  eigen value  $\mu_1=0$ corresponds to  the perturbation parallel to the synchronization manifold, while the other eigen values correspond to the perturbations  transverse to the synchronization manifold. The transverse eigenmodes should be damped out to have a stable synchronization manifold.  By substituting the eigen values  in  Eq.~(\ref{msf1})-(\ref{msf2}) and finding the largest  Lyapunov  exponents, one can distinguish stable regions of synchronized and desynchronized states.  Whenever the largest  Lyapunov  exponents acquire negative values the   synchronized manifold is stable. Figs.~\ref{fig12}(a)-(b) elucidate that the value of $\Lambda_{max}$ obtained from the variational equations distinguish the synchronized and desynchronized regimes for fixed values of microwave current and microwave field. In order to verify the above analytical results we have plotted the standard deviation $\sigma=<\sqrt{\frac{1}{N}\sum_{j}(m^{x}_{j}-\bar{M})^{2}}>_{t}$, $\bar{M}=\frac{1}{N}\sum_{j}m^{x}_{j}$, of the magnetization vector component $m^{x}_{j}$ by varying the coupling strength in Figs.~\ref{fig12}(c)-(d) for the same values of microwave current and microwave field for $N=10,50$ and 100.  Here,  we find that our analytical results match with our numerically obtained results and stability of synchronized states independent of $N$. Also, one can note that when the value of $\Lambda_{max}$ crosses the zero value along the $x$-axis when changing from from positive to negative values as $\epsilon$ increases, the value of $\sigma$ takes zero indicating the   onset of synchronization among the STNOs.

\section{Conclusion}
In this paper, we have first studied the locking characteristics of STNOs coupled  indirectly through a common microwave current or a common microwave field. We find that when the external frequency is higher than that of the natural frequency of oscillation  the system can show second harmonic locking and locking behaviour takes place for lesser values of microwave current and field when compared with first harmonic locking.

 We have established several integer harmonic lockings in the case of ac current injection.  However, in the case of microwave field,  in addition to integer harmonic lockings, several fractional lockings  have been identified.  The occurrence of fractional lockings in the Arnold tongue diagram  is found to be in good agreement with experimental results also
\cite{tabor:10,tabor1:10}.   We have also studied the locking behaviour of STNOs with respect to anisotropy. From this, we have demonstrated the possibility of locking only exists for large number of STNOs through second harmonic locking.  We have also studied the locking behavior of STNOs by introducing a self-delay feedback, where it is shown that the delay feedback oscillator does not get synchronized with non-delayed feedback oscillator  when it exceeds  certain critical value in the first harmonic locking. But, the delay feedback oscillator always gets synchronized with respect to delay factor as in the case of second harmonic locking.

Further, we also examined the frequency locking characteristics of two STNOs coupled through current injection. Our results  show that current injection between two oscillators with appropriate strength  makes it possible to get locking at  rather low values of microwave current and field. It also shows that several fractional lockings also exist in the case of microwave field  when compared with microwave current.  Finally, we have also carried out stability analysis of synchronization of large arrays of STNOs with current coupling by using master stability function formalism. We also find that our analytical results from master stability function formalism match with numerically obtained results. The results of our studies suggest a tunable microwave field which is a promising configuration for STNO applications in the field of communication systems.

\section*{Acknowledgments}
B. Subash wishes to thank the Department of Science and Technology, Government of India for financial support in the form of a National Postdoctoral Fellowship(PDF/2016/000686). The work of V.K.C. forms part of a research project sponsored by INSA Young Scientist Project under Grant No. SP/YSP/96/2014. M. L wishes to thank the Department of Science and Technology for the award of a SERB Distinguished Fellowship under Grant No.SERB/F/6717/2017-18. His work also forms part of a CSIR Project (03/1331/15/EMR -II).

\end{document}